\documentclass[letter,traditabstract]{aa}  

\usepackage{graphicx}
\usepackage{natbib}
\usepackage{lscape}
\usepackage{longtable}
\usepackage{txfonts}

\usepackage{float}

\newcommand{\degree}{\ensuremath{^\circ}}

\begin{document}

   \title{Gaia DR2 view of the Lupus V-VI clouds: the candidate diskless young stellar objects are mainly background contaminants}

\titlerunning{Lupus V-VI with Gaia DR2}
\authorrunning{Manara et al.}

   \author{C.F. Manara \inst{1}\fnmsep\thanks{ESO Fellow} 
          \and
T. Prusti\inst{2}
          \and
F. Comeron\inst{1}
          \and
R. Mor\inst{3}
          \and
J.M. Alcal\'a\inst{4}
          \and
T. Antoja\inst{3}
          \and
S. Facchini\inst{5}
          \and
D. Fedele\inst{6}
          \and
A. Frasca\inst{7}
          \and
T. Jerabkova\inst{1}
          \and
G. Rosotti\inst{8}
          \and
L. Spezzi\inst{9}
          \and
L. Spina\inst{10}
          }

   \institute{European Southern Observatory, Karl-Schwarzschild-Strasse 2, 85748 Garching bei M\"unchen, Germany\\
              \email{cmanara@eso.org}
\and
Scientific Support Office, Directorate of Science, European Space Research and Technology Centre (ESA/ESTEC), Keplerlaan 1, 2201 AZ Noordwijk, The Netherlands
\and
Institut de Ci\`encies del Cosmos, Universitat de Barcelona (IEEC-UB), Mart\'i i Franqu\`es 1, E-08028 Barcelona, Spain
\and
INAF-Osservatorio Astronomico di Capodimonte, via Moiariello 16, 80131 Napoli, Italy
\and
Max-Planck-Institut f\"ur Extraterrestrische Physik, Giessenbachstrasse 1, 85748 Garching bei M\"unchen, Germany
\and
INAF-Osservatorio Astrofisico di Arcetri, L.go E. Fermi 5, 50125 Firenze, Italy
\and 
INAF-Osservatorio Astrofisico di Catania, via S. Sofia, 78, 95123 Catania, Italy
\and
Institute of Astronomy, University of Cambridge, Madingley Road, Cambridge CB3 0HA, UK
\and
European Organisation for the Exploitation of Meteorological Satellites (Eumetsat),  Eumetsat Allee 1, 64295 Darmstadt, Germany
\and
Monash Centre for Astrophysics, School of Physics and Astronomy, Monash University, VIC 3800, Australia
}
             
   \date{Received May 7, 2018; accepted June 11, 2018}

 
  \abstract
  {Extensive surveys of star-forming regions with \textit{Spitzer} have revealed populations of disk-bearing young stellar objects. These have provided crucial constraints, such as the timescale of dispersal of protoplanetary disks, obtained by carefully combining infrared data with spectroscopic or X-ray data. While observations in various regions agree with the general trend of decreasing disk fraction with age, the Lupus V and VI regions appeared to have been at odds, having an extremely low disk fraction. Here we show, using the recent Gaia data release 2 (DR2), that these extremely low disk fractions are actually due to a very high contamination by background giants. Out of the 83 candidate young stellar objects (YSOs) in these clouds observed by Gaia, only five have distances of $\sim$150 pc, similar to YSOs  in the other Lupus clouds, and have similar proper motions to other members in this star-forming complex. Of these five targets, four have optically thick (Class~II) disks. On the one hand, this result resolves the conundrum of the puzzling low disk fraction in these clouds, while, on the other hand, it further clarifies the need to confirm the \textit{Spitzer} selected diskless population with other tracers, especially in regions at low galactic latitude like Lupus V and VI. The use of Gaia astrometry is now an independent and reliable way to further assess the membership of candidate YSOs in these, and potentially other, star-forming regions.  }


   \maketitle
%

\section{Introduction}
The timescale on which protoplanetary disks evolve is a key constraint on planet formation and disk evolution models. Observations show that the fraction of young stellar objects (YSOs) surrounded by optically thick disks decreases exponentially with time with a typical timescale of $\sim$2-3 Myr \citep[e.g.,][]{haisch01,hernandez07,fedele10}. 

In this context, the very low disk fraction of about 15\% derived with \textit{Spitzer} in the Lupus V and VI clouds by \citet{spezzi11}, significantly smaller than the value of about 50\% \citep{merin08} for the nearby 2-3-Myr-old  \citep{alcala14,frasca17}  Lupus I and III clouds, appears to be at odds with the general trend. Indeed, if the age of these clouds is compatible with that of the other parts of the Lupus complex, and the candidate diskless (Class~III) YSOs are confirmed to be members of the cloud population,  other processes would be required to explain the particularly fast dispersal of disks \citep{spezzi11}, such as external photoevaporation or higher binarity fraction \citep[e.g.,][]{facchini16,rosotti18}. Since the Class~III population of \textit{Spitzer}-selected candidate YSOs is known to be contaminated by background objects with contaminant fractions between 25 and 90\% \citep{dunham15}, independent data must be used to confirm the YSO status of the candidates. An analysis of 31 of these YSO candidates in these two clouds, conducted by \citet{romero12} using optical spectroscopy, showed that this sub-sample of the population is made up of background contaminants. In a re-analysis of the same \textit{Spitzer} data by \citet{dunham15}, 11 objects in Lupus~V and 17 in Lupus~VI were classified as likely AGB contaminants. However, the majority of the Spitzer candidates are still to date not confirmed as legitimate YSOs.

Here we adopt another powerful method, independent of the one used by \citet{romero12}, to discriminate members of the star-forming region from background contaminants. We explore the astrometric parameters for the whole YSO candidate population of the Lupus V and VI clouds from the recent Gaia \citep{gaia} data release 2 \citep[DR2,][]{gaiadr2,gaiadr2astrometry}. Using their parallax, as well as their proper motions and location on the color-magnitude diagram, we can determine the real nature of the candidate YSOs in these clouds to further constrain how protoplanetary disks evolve.


\section{Sample and data}\label{sect::data}

The Lupus cloud complex is a well-known nearby low-mass star-forming complex located at a distance of $\sim$150-200 pc \citep{comeron08,gaiadr2}. It is composed of several molecular clouds, whose density peaks, or clumps, are usually referred to as distinct clouds. In particular, the largest CO molecular cloud of the complex contains the Lupus III, IV, V, and VI clouds \citep{tachihara01}.
Lupus was one of the targets of the {\it Spitzer} Legacy surveys {\it From Molecular Cores to Planet-forming Disk} \citep[c2d, PI: N. Evans;][]{evans09} and {\it Gould's Belt} \citep[GB, PI: L. Allen;][]{dunham15}. Among the various regions targeted in these surveys, the list of YSO candidates from which \citet{spezzi11} derived the comparably low disk fraction includes 43 and 45 objects in the Lupus~V and VI clouds, respectively. The analysis of the same dataset by \citet{dunham15} confirmed the candidate YSO status for 32 and 28 of these targets, respectively, while suggesting that the remaining objects are likely AGB contaminants. Furthermore, they suggest that the contaminant fraction among the Class~III YSO candidates could be as high as 90\%. In this work, we use the complete initial list by \citet{spezzi11} for the analysis.

   \begin{figure}
   \centering
  \includegraphics[scale=0.45]{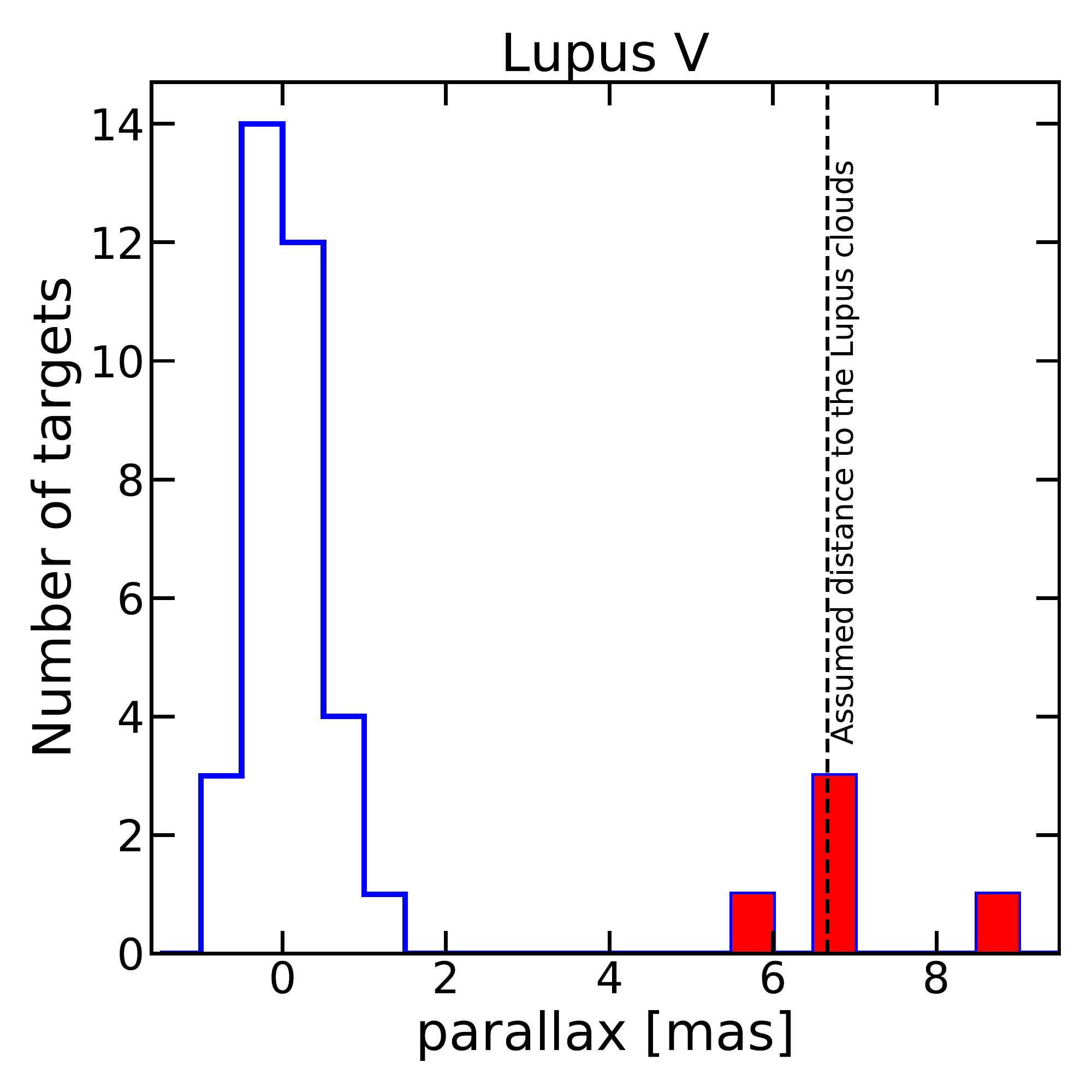}
  \includegraphics[scale=0.45]{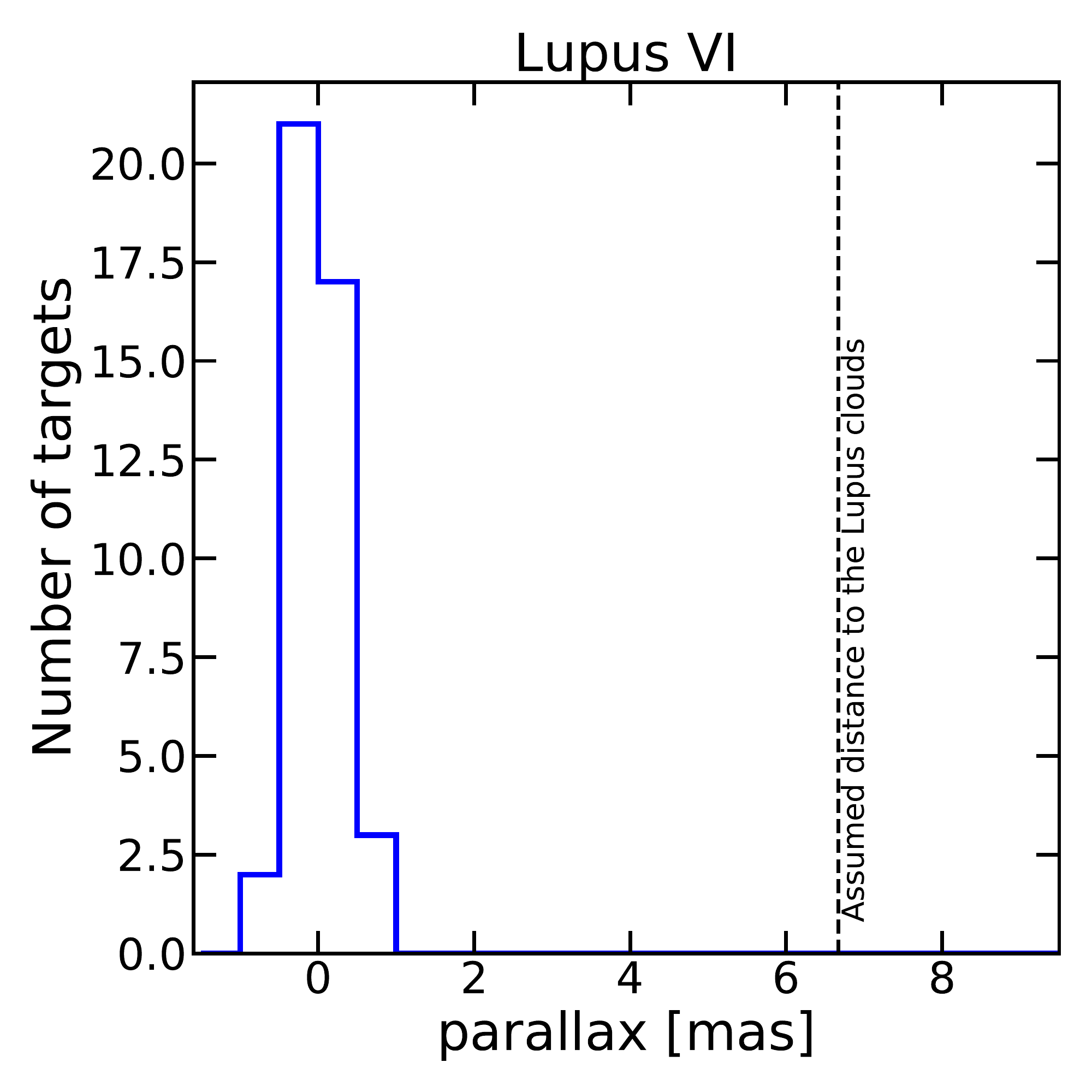}
   \caption{Histogram of measured parallaxes for the candidate YSOs proposed by \citet{spezzi11} in the Lupus~V (top) and Lupus~VI (bottom) clouds. Highlighted in red are the objects with parallax$>$5 mas. These are five objects in the Lupus~V cloud, which are the only real YSOs among the \textit{Spitzer}-selected candidates. }
              \label{fig::parhist}%
    \end{figure}

\subsection{Gaia data collection}
We have downloaded the data from the Gaia archive using the ADQL queries reported in Appendix~\ref{adql_queries}. We first select all objects within 243.5\degree $<$ ra $<$ 246.9\degree \ and  -38.5\degree $<$ dec $<$ -36.3\degree \ (Lupus~V) and 244.4\degree $<$ ra $<$ 247.8\degree \ and  -42.1\degree $<$ dec $<$ -38.7\degree \ (Lupus~VI). These searches find, in the Gaia DR2 catalog, a total of 663071 targets in the Lupus V region of the sky and 1517387 in the Lupus VI field. We then use the cross-match of the Gaia catalog with the 2MASS catalog provided by the Gaia consortium within the Gaia DR2 to find the best match within radii of 1\arcsec (see Appendix~\ref{adql_queries}), and find, in the same regions of the sky, 188053 matches for the Lupus V cloud region, and 440565 for the Lupus VI region. 

We then proceed by matching the catalog of candidate YSOs by \citet{spezzi11} with the Gaia DR2 catalog. We check that the matches between Gaia and 
the catalog by \citet{spezzi11} are always with angular separations $\lesssim 0.3$\arcsec. We find a total of 39 out of 43 matches for the Lupus~V region, and 44 out of 45 for the Lupus~VI region. The objects with no Gaia counterpart are the faintest in the samples ($J\sim13-16$ mag, $K\sim15$ mag). The one candidate YSO in Lupus~VI with no match is classified by \citet{spezzi11} as \textit{Flat} source, while three of the four targets with no match in Lupus~V are classified as Class~III, and the last one as Class~II based on their \textit{Spitzer} infrared(IR)-excess up to 24 $\mu$m.

   \begin{figure}
   \centering
  \includegraphics[scale=0.45]{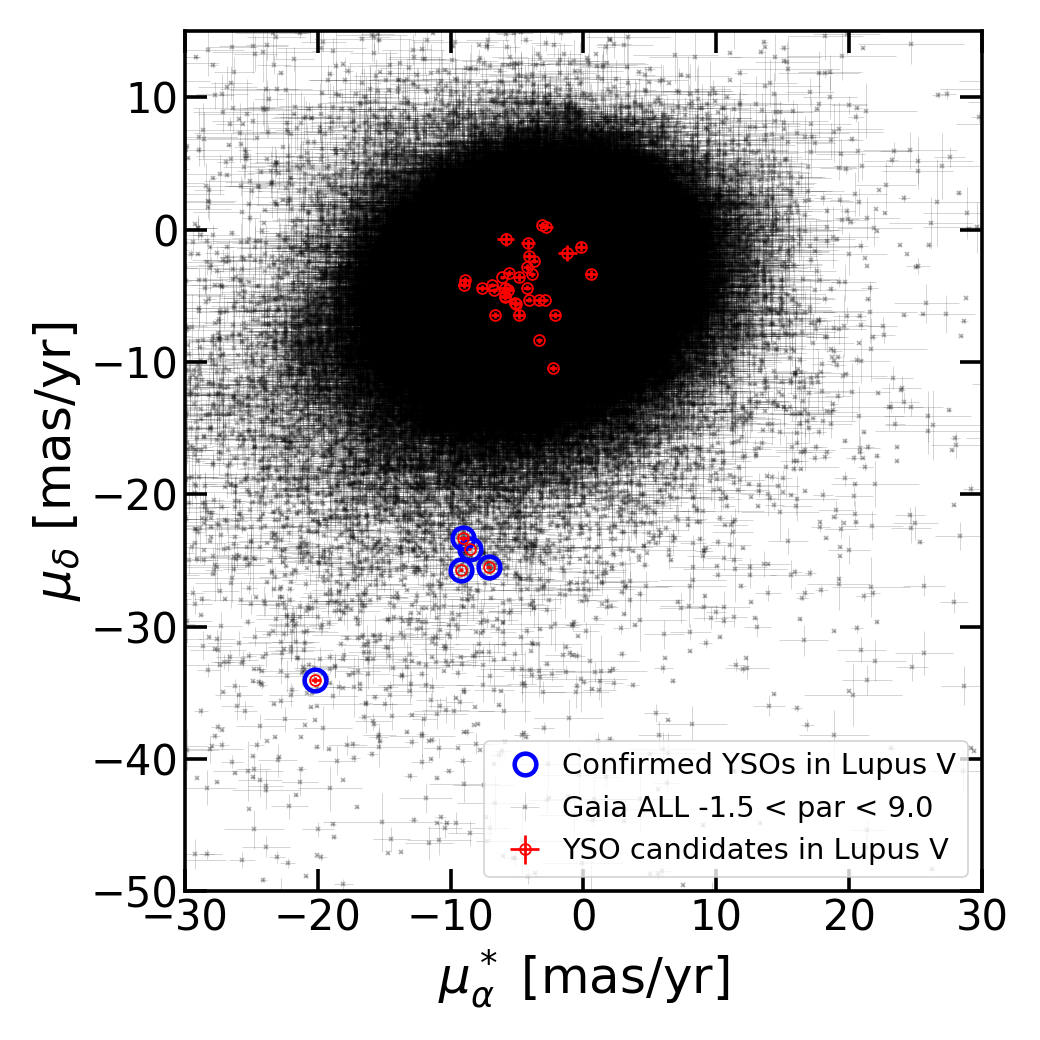}
   \caption{Proper motion of the Gaia targets in the field of view of the Lupus~V region, highlighting the candidate young stellar objects with red circles, and the five confirmed young stars with cyan circles. }
              \label{fig::propmot}%
    \end{figure}
%

   \begin{figure}
   \centering
  \includegraphics[scale=0.45]{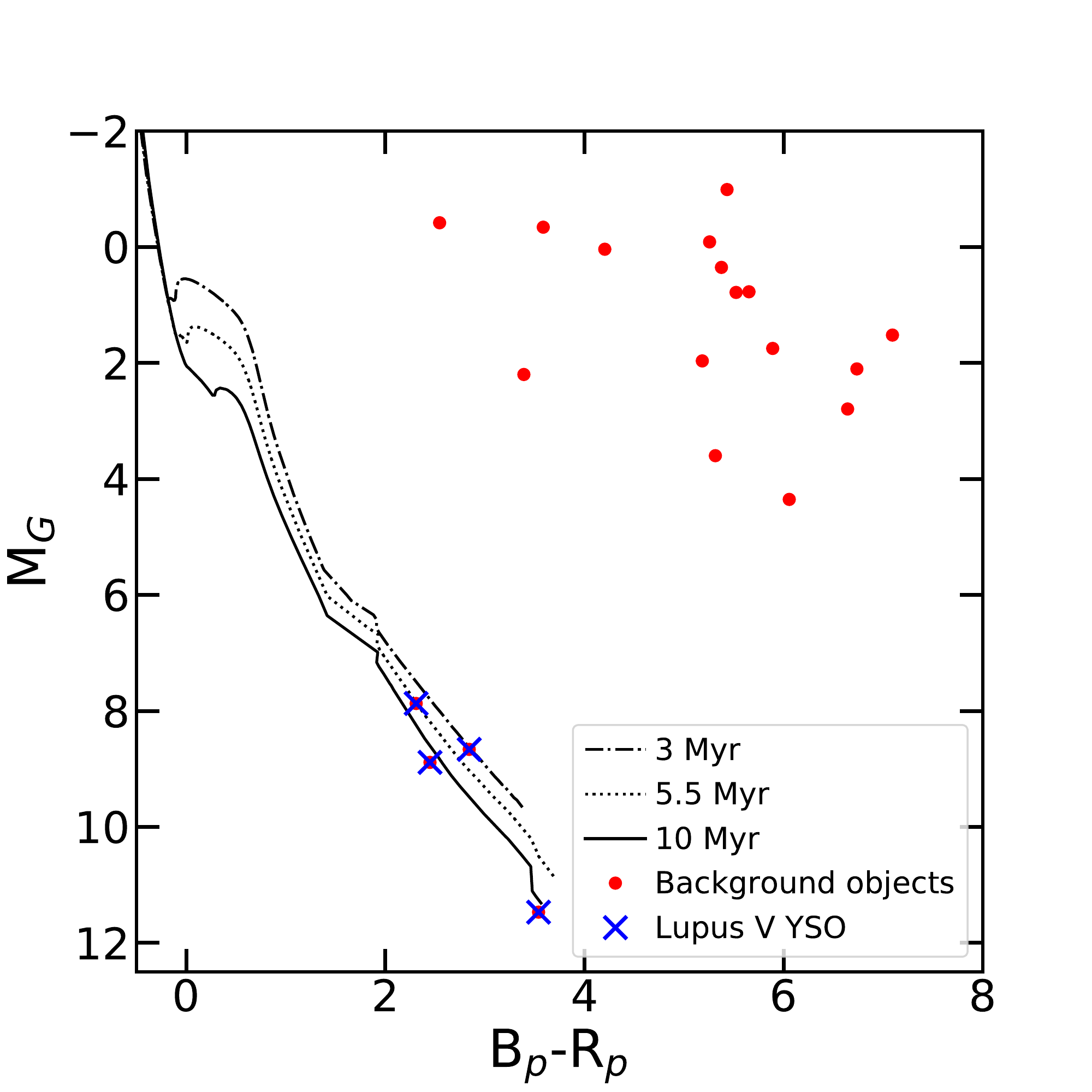}
  \includegraphics[scale=0.45]{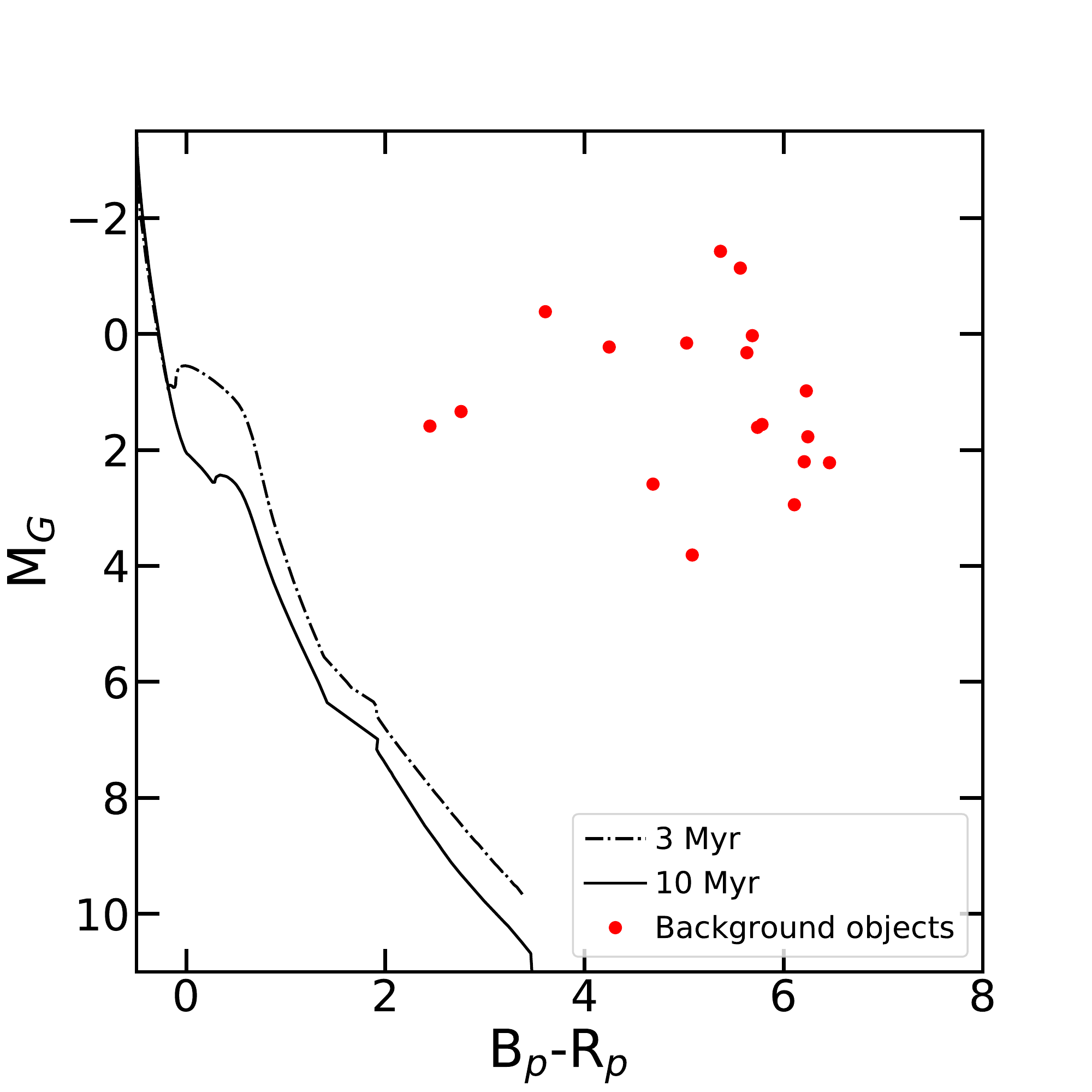}
   \caption{Absolute magnitude vs color for the candidate YSOs in the Lupus~V (top) and Lupus~VI (bottom) regions with positive parallax and available Gaia magnitudes. Only the four confirmed young objects in Lupus~V with available Gaia magnitudes have color and magnitudes compatible with age$<$10 Myr, according to the overplotted extinction-free isochrones.}
              \label{fig::cmd}%
    \end{figure}
%


\section{Analysis of Gaia data}\label{sect::analysis}

The first indication that an object is a member of the Lupus complex is found by examining its parallax. The distribution of parallaxes for the YSO candidates in these two clouds is shown in Fig.~\ref{fig::parhist}. Of these targets, only five objects in Lupus V have parallaxes larger than 5 mas, corresponding to distances smaller than 200 pc, while all the other objects have parallax $\lesssim$ 2 mas, implying distances larger than 500 pc. No candidate YSOs in the Lupus VI region have parallax $>$ 1 mas. This information already suggests that most of the candidate YSOs selected by \citet{spezzi11} are, in reality, background objects. 

We also explored whether there is any dependence of the measured parallaxes of the YSO candidates with their position in the sky, and we find that the five objects with parallaxes larger than 5 mas are all in low-extinction parts of the cloud, mainly on the west side, in the direction of the Lupus~III cloud (see Appendix~\ref{app::par_sky}). 
We also find that four of the YSO candidates with parallaxes larger than 5 mas, and therefore distances smaller than 200 pc, are classified as Class II by \citet{spezzi11}, and that the remaining one is classified as Class~III. This classification is confirmed also by \citet{dunham15}, and none of these five targets are flagged as likely being AGB contaminants. These five objects have a weighted mean parallax of 6.662$\pm$0.013 mas, corresponding to a mean distance of $\sim$150 pc. If the Lupus clouds are at 150 pc \citep[e.g.,][]{comeron08}, then three of these targets are in the cloud, while one is in front of it and one behind it.

To further investigate whether the closer YSO candidates are different from the background objects, we show in Fig.~\ref{fig::propmot} the proper motion for the YSO candidates in the Lupus~V cloud. Four out of the five objects with parallaxes larger than 5 mas are all clustered at $\mu_\alpha^* \sim -9$ mas/yr, $\mu_\delta \sim -25$ mas/yr, and the last one has $\mu_\alpha^* \sim -20$ mas/yr, $\mu_\delta \sim -34$ mas/yr. These are very different values from those of the bulk of the candidate YSOs, which are found to have $\mu_\alpha^* \sim -2$ mas/yr, $\mu_\delta \sim -2$ mas/yr, the same parameter space as most of the Gaia targets in the field of view, which are mainly background objects. The difference in the proper motion space of these closer YSO candidates suggests that they are a homogeneous group that is kinematically distinct from the more distant objects. This confirms that they most likely belong to the cloud population. When comparing the proper motion values for the five close-by targets with those of other confirmed YSOs in the Lupus complex \citep{lopez-marti11}, these are found in very good agreement with those with higher proper motion in this region. This confirms these five objects as members of the Lupus star-forming region.

Finally, we collect the $G$, $B_P$, $R_P$ magnitudes for all the targets included in the Gaia DR2 catalog. Following \citet{gaiadr2_hdr}, we convert the $G$ magnitudes into absolute magnitudes as $M_G =
G + 5 + 5 log_{10}(\omega/1000)$, with $\omega$ being the parallax in milliarcseconds. Since this works only for targets with measured magnitudes and positive parallaxes, the absolute magnitude versus color plots in Fig.~\ref{fig::cmd} show only a fraction of the YSO candidates. Nevertheless, the comparison of the position of the YSO candidates on these color-magnitude diagrams, especially when compared with the isochrones by \citet{parsec}, shows that four out of five of the nearby targets in Lupus~V are indeed young objects with isochronal ages between 3 and 10 Myr, while all the other YSO candidates are instead located in the part of the color-magnitude diagram compatible with the locus of giants. No $B_P$ and $R_P$ magnitudes are available for one of the five bona fide YSOs in Lupus~V. Neglecting the effect of interstellar extinction has a minor impact, as these young objects are all located in regions of the clouds where $E(B-V)$ is smaller than 1.4 mag (see Appendix~\ref{app::par_sky}), and this would just change the isochronal age of the objects by $\lesssim$1 Myr.

\begin{table*} 
\caption{\label{tab::YSOs} Properties of the confirmed YSOs in the Lupus~V region.}
\centering 
\begin{tabular}{l*{6}{c}}
\hline
2MASS J & Gaia DR2 ID & ra & dec & parallax & \\
 & & [deg] & [deg] & [mas] & \\
\hline
16191403-3747280 & 6021420630046381440 & 244.80843901 & $-$37.79119223 & 6.568 $\pm$ 0.042 & \\
16172485-3657405 & 6021662385163162240 & 244.35352824 & $-$36.96138759 & 6.760 $\pm$  0.067 &    \\
16172475-3657332 & 6021662385163163648 & 244.35306973 & $-$36.95935461 & 6.588 $\pm$ 0.082 & ... \\
16171811-3646306 & 6021805356032645504 & 244.32543416 & $-$36.77529065 & 5.957 $\pm$ 0.179 &    \\
16192684-3651235 & 6021745462701109376 & 244.86172975 & $-$36.85671498 & 8.910 $\pm$ 0.187 &  \\
\end{tabular}
\begin{tabular}{c*{5}{c}}
\hline
$\mu_\alpha^*$  & $\mu_\delta$ & $G$ & $B_P$ & $R_P$ & Lada\\
\hbox{}[mas/yr]& [mas/yr] & [mag] & [mag] & [mag] & Class \\
\hline
$-$8.56 $\pm$  0.10 & $-$24.10 $\pm$ 0.06 & 13.78 & 14.93& 12.62 & II \\
$-$9.21 $\pm$  0.14 &  $-$25.68 $\pm$ 0.08 &  14.51 & 16.12 & 13.28 & III \\
 $-$7.14 $\pm$ 0.18 &  $-$25.50 $\pm$ 0.10 & 14.79 & 16.06 & 13.61 & II \\
$-$9.03 $\pm$ 0.36 &  $-$23.32  $\pm$ 0.20 & 17.64 & ... & ... & II \\
$-$20.18 $\pm$ 0.33 &  $-$34.04 v 0.24 & 16.72 & 18.87 & 15.33 & II \\
\hline
\end{tabular}
\tablefoot{Reference epoch: 2015.5. Lada classes are from \citet{spezzi11}.}
\end{table*}


\section{Discussion}

Only five out of 83 candidate YSOs in the Lupus V and VI clouds are confirmed YSOs. 
These five targets have similar proper motions to other YSOs in the Lupus complex and four of them have IR excess typical of Class~II objects. They could either be a different population with respect to the nearby Lupus~III cloud, or members of the latter. 
If they are an independent population, then the disk fraction found here could seem too high for their age (some millions of years). However, the age estimate is subject to large uncertainties \citep[e.g.,][]{soderblom14} and we cannot asses whether we have a complete census of this putative YSO population of Lupus~V. 
It is to be expected that a larger population of diskless YSOs is also present in this cloud. The location of the clouds being to the east supports instead the hypothesis that these objects are members of Lupus~III, and they are located off-cloud, like the YSOs studied by \citet{comeron09}.

The lack of confirmed YSOs in Lupus~VI and the small number of confirmed YSOs in Lupus~V could imply that either star formation has not started in these clouds, like in the Musca cloud north of the Chamaeleon complex \citep{musca88,cox16}, or that these clouds are actually relatively sterile. \citet{spezzi11} discussed the fact that the extinction maps for these regions peak at $A_V\lesssim$ 6 mag, which is in the range of values where star formation is highly inefficient \citep{lada13}. Therefore, it is indeed possible that these clouds are not able to form stars now.

This work advocates caution when dealing with the conclusions derived from disk statistics based on large numbers of \textit{Spitzer}-selected Class~III objects, which are prone to contamination by unrelated types of objects, as extensively discussed also by \citet{dunham15}. This effect is lower in regions dominated by a high density of Class~II YSOs, which are selected with a much higher level of confidence by \textit{Spitzer} color criteria. Indeed, the spectroscopic follow-up of \textit{Spitzer} surveys carried out by, for example, \citet{oliveira09} in the Serpens region, or \citet{alcala14,alcala17} in the Lupus I, III, and IV clouds, looking for the presence of lithium, H$\alpha$, and several other indicators of youth, have shown that about 95\% of the \textit{Spitzer} Class~II candidates are confirmed as legitimate YSOs. Only a handful of contaminant background giants are found in these regions \citep[see also][]{dunham15,frasca17}. In the case of Lupus~V and VI, however, the candidate YSOs were mainly Class~III, and also \citet{oliveira09}, as well as \citet{romero12} and \citet{dunham15}, have shown that the fraction of contaminants can be larger than 50\% for diskless stars. Furthermore, the low location close to the galactic plane of the Lupus~V cloud ($b\sim8.8\degree$), and the even lower location of the Lupus~VI cloud ($b\sim6.3\degree$), implies that contamination from background giants can be much higher than in star-forming regions well away from the galactic plane, as also pointed out by \citet{dunham15}. We verified the number of background giants expected in the Gaia catalog in these regions of the sky by exploring the content of the Gaia Universe Model Snapshot \citep[GUMS;][]{robin12}. We selected all the stars with $G<$21.5 mag from the same region of the sky as the one that we queried from the Gaia archive (see Sect.~\ref{sect::data}). The number of selected stars is of the same order of magnitude as those found in the Gaia archive. By exploring the color-magnitude diagram for these targets, we found that there are $\sim 3\times 10^4$ and $\sim 10^5$ background giants in these fields, of which the majority ($\sim 65-75\%$) are red giants. Furthermore, the number of background contaminants found here is of the order of the estimated red giants of spectral types M7, M8, M9, C-type and S-type in the Lupus V and Lupus VI fields predicted by GUMS, that is, 50 and 150 in the Lupus~V and Lupus~VI fields, respectively. These are the reddest giants, and are the likely contaminants of the \textit{Spitzer} YSO selection criteria. 
This means that the number of background giants erroneously classified as Class~III YSO candidates in previous \textit{Spitzer} surveys of these regions is just a tiny fraction of the total number of background objects. This implies, on the one hand, that the high fraction of contaminants present in these samples is in line with the number of the reddest giants, while there are orders of magnitude more giants in the field. On the other hand, this also implies that not all of the background objects are classified with \textit{Spitzer} as Class~III YSO candidates, as many are correctly excluded by the color-color selection criteria.

On top of spectroscopy and the detection of X-ray emission from these YSO candidates, we have shown here that the analysis of the astrometric properties of the targets obtained with Gaia is a powerful tool to discriminate between bona fide YSOs and contaminants.

%

\section{Conclusions}

Here, we investigate the astrometric properties of the \textit{Spitzer} selected candidate YSOs in the Lupus V and VI clouds and find that only five targets have parallaxes larger than 5 mas, proper motions compatible with other YSOs in the Lupus complex, and occupy the right region of the color-magnitude diagram to be considered YSOs. Of these five objects, four are shown to have optically thick disks based on their \textit{Spitzer} colors. This reconciles the disk fraction in Lupus V-VI with the other regions in the Lupus clouds complex, without the need to invoke an exceptionally short disk lifetime. These targets are located at $d\sim$150 pc and have ages between 3 and 10 Myr. These objects are either members of the Lupus~V region, or off-cloud YSOs of Lupus~III. In the former case, it is possible that several Class~III objects in this region have not yet been detected by previous surveys. Furthermore, these two clouds, to our knowledge only hosting a total of five YSOs, are possibly sterile or have not yet begun  forming stars.

The remaining 78 YSO candidates in these clouds present in the Gaia DR2 catalog are not YSOs, but background objects, mainly red giants. Previous claims of an anomalously low disk fraction can therefore be explained by the very low star formation activity in Lupus V and VI, the low reliability of selection criteria based on \textit{Spitzer} colors when applied to Class~III sources, and by the low galactic latitude of both clouds, which increases the areal density of potential contaminants.

The work presented here highlights the power of the astrometric information supplied by Gaia as a new fundamental tool to obtain reliable samples of bona fide members of nearby star-forming regions.

\begin{acknowledgements}
We thank the anonymous referee for the constructive report which helped to improve the quality of the work.
      This work has made use of data from the European Space Agency (ESA) mission
{\it Gaia} (\url{https://www.cosmos.esa.int/gaia}), processed by the {\it Gaia}
Data Processing and Analysis Consortium (DPAC,
\url{https://www.cosmos.esa.int/web/gaia/dpac/consortium}). Funding for the DPAC
has been provided by national institutions, in particular the institutions
participating in the {\it Gaia} Multilateral Agreement.  We made use of the Python packages numpy, astropy, aplpy, and matplotlib for the analysis.
CFM acknowledges support with an ESO Fellowship. 
JMA acknowledge financial support from the project PRIN-INAF 2016
   The Cradle of Life - GENESIS-SKA (General Conditions in Early
   Planetary Systems for the rise of life with SKA). D.F. and acknowledge support from the Italian Ministry of Education, Universities and Research, project SIR (RBSI14ZRH).
This work has been supported by the DISCSIM project, grant
agreement 341137 funded by the European Research Council
under ERC-2013-ADG
\end{acknowledgements}


\appendix
\onecolumn

\section{ADQL queries}\label{adql_queries}

Part of sky of Lupus V:

\begin{verbatim}
SELECT
        gaia.solution_id, gaia.designation, gaia.source_id, gaia.ref_epoch, gaia.ra, gaia.ra_error, gaia.dec, 
    gaia.dec_error, gaia.parallax, gaia.parallax_error, gaia.parallax_over_error, gaia.pmra, gaia.pmra_error,
    gaia.pmdec, gaia.pmdec_error, gaia.astrometric_gof_al, gaia.astrometric_params_solved, gaia.phot_g_mean_mag,
    gaia.phot_bp_mean_mag, gaia.phot_rp_mean_mag, gaia.radial_velocity, gaia.radial_velocity_error 
    FROM gaiadr2.gaia_source as gaia
    WHERE
        gaia.dec > -38.5 AND gaia.dec < -36.3
        AND gaia.ra > 243.5 AND gaia.ra < 246.9
\end{verbatim}

Part of sky of Lupus VI:
\begin{verbatim}
SELECT
        gaia.solution_id, gaia.designation, gaia.source_id, gaia.ref_epoch, gaia.ra, gaia.ra_error, gaia.dec, 
    gaia.dec_error, gaia.parallax, gaia.parallax_error, gaia.parallax_over_error, gaia.pmra, gaia.pmra_error,
    gaia.pmdec, gaia.pmdec_error, gaia.astrometric_gof_al, gaia.astrometric_params_solved, gaia.phot_g_mean_mag,
    gaia.phot_bp_mean_mag, gaia.phot_rp_mean_mag, gaia.radial_velocity, gaia.radial_velocity_error 
    FROM gaiadr2.gaia_source as gaia
    WHERE
        gaia.dec > -42.1 AND gaia.dec < -38.7
        AND gaia.ra > 244.4 AND gaia.ra < 247.8
\end{verbatim}

And also the following data cross-matched with 2MASS:

\begin{verbatim}
SELECT
        *
    FROM gaiadr2.gaia_source as gaia
    INNER JOIN gaiadr2.tmass_best_neighbour as xmatch ON gaia.source_id = xmatch.source_id
    INNER JOIN gaiadr1.tmass_original_valid as tmas ON tmas.tmass_oid = xmatch.tmass_oid
    WHERE
        gaia.dec > -38.5 AND gaia.dec < -36.3
        AND gaia.ra > 243.5 AND gaia.ra < 246.9
        AND xmatch.angular_distance < 1.0
\end{verbatim}

\begin{verbatim}
SELECT
        *
    FROM gaiadr2.gaia_source as gaia
    INNER JOIN gaiadr2.tmass_best_neighbour as xmatch ON gaia.source_id = xmatch.source_id
    INNER JOIN gaiadr1.tmass_original_valid as tmas ON tmas.tmass_oid = xmatch.tmass_oid
    WHERE
        gaia.dec > -42.1 AND gaia.dec < -38.7
        AND gaia.ra > 244.4 AND gaia.ra < 247.8
        AND xmatch.angular_distance < 1.0
\end{verbatim}

\section{Parallaxes as a function of position in the sky}\label{app::par_sky}

\begin{figure}
   \centering
  \includegraphics[scale=0.45]{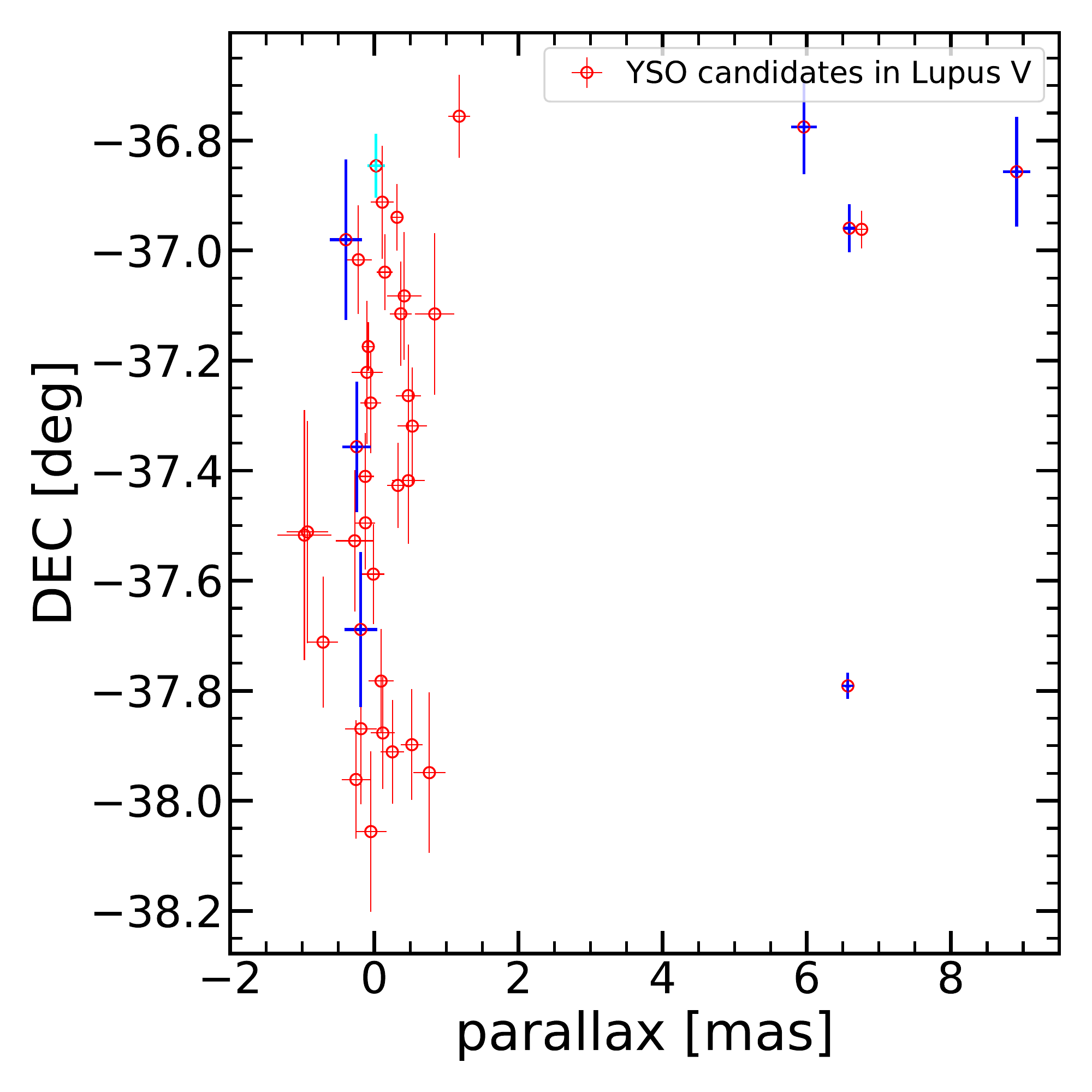}
  \includegraphics[scale=0.45]{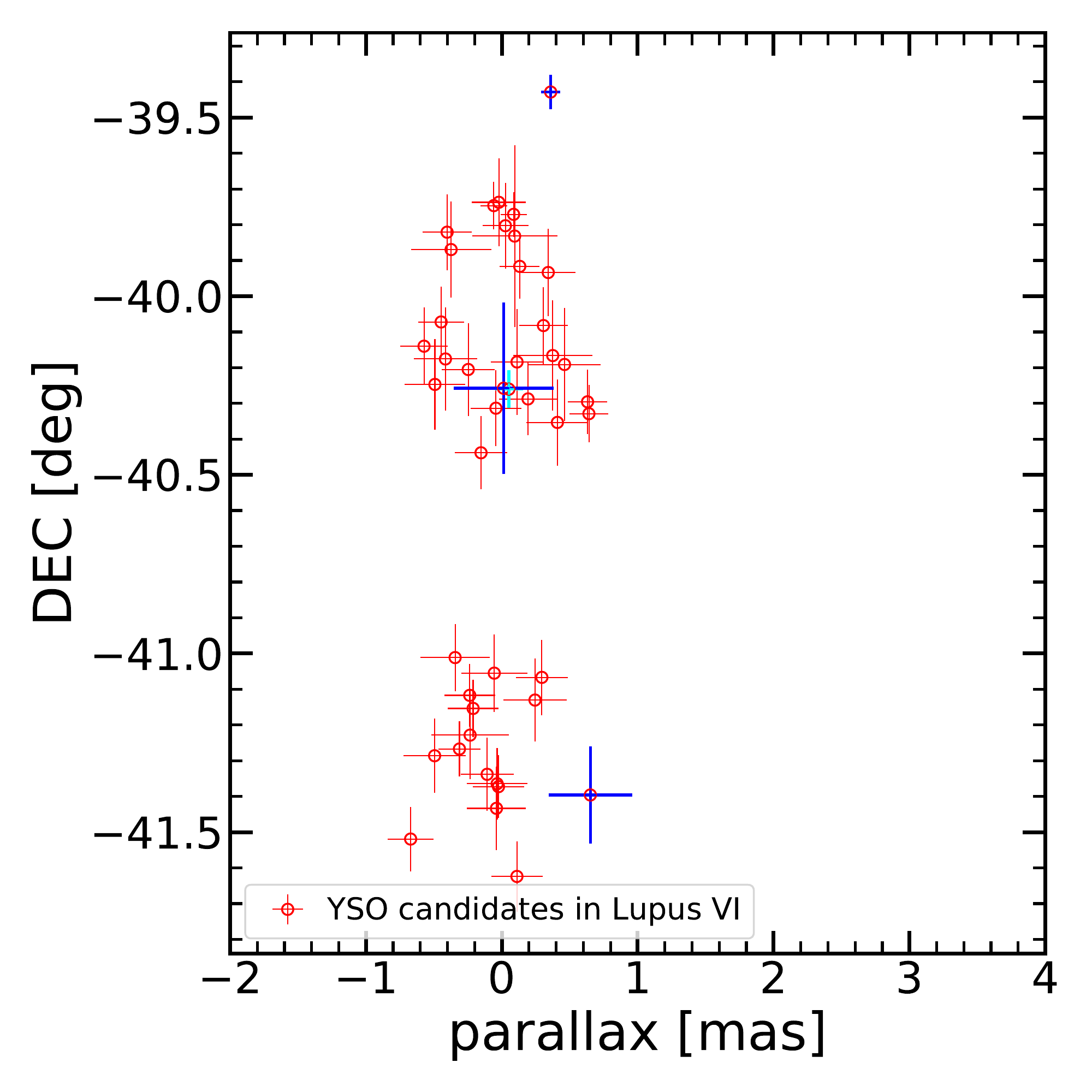}
   \caption{Parallaxes of candidate YSOs as a function of declination for objects in the Lupus~V (left) and Lupus~VI (right) region of the sky. Blue crosses indicate candidate Class~II objects, cyan crosses candidate transition disk objects. }
              \label{fig::pardec}%
    \end{figure}

\begin{figure}
   \centering
  \includegraphics[scale=0.45]{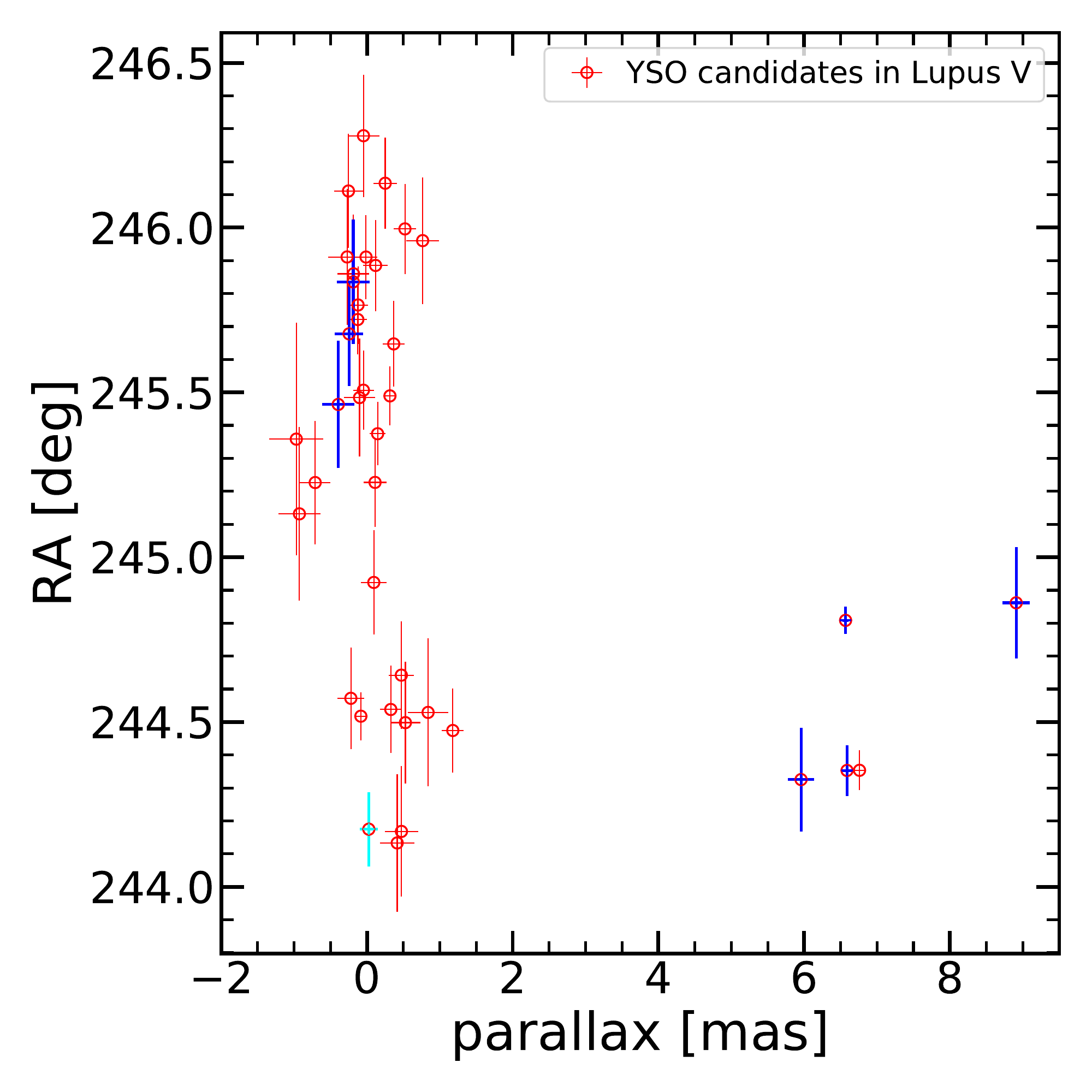}
  \includegraphics[scale=0.45]{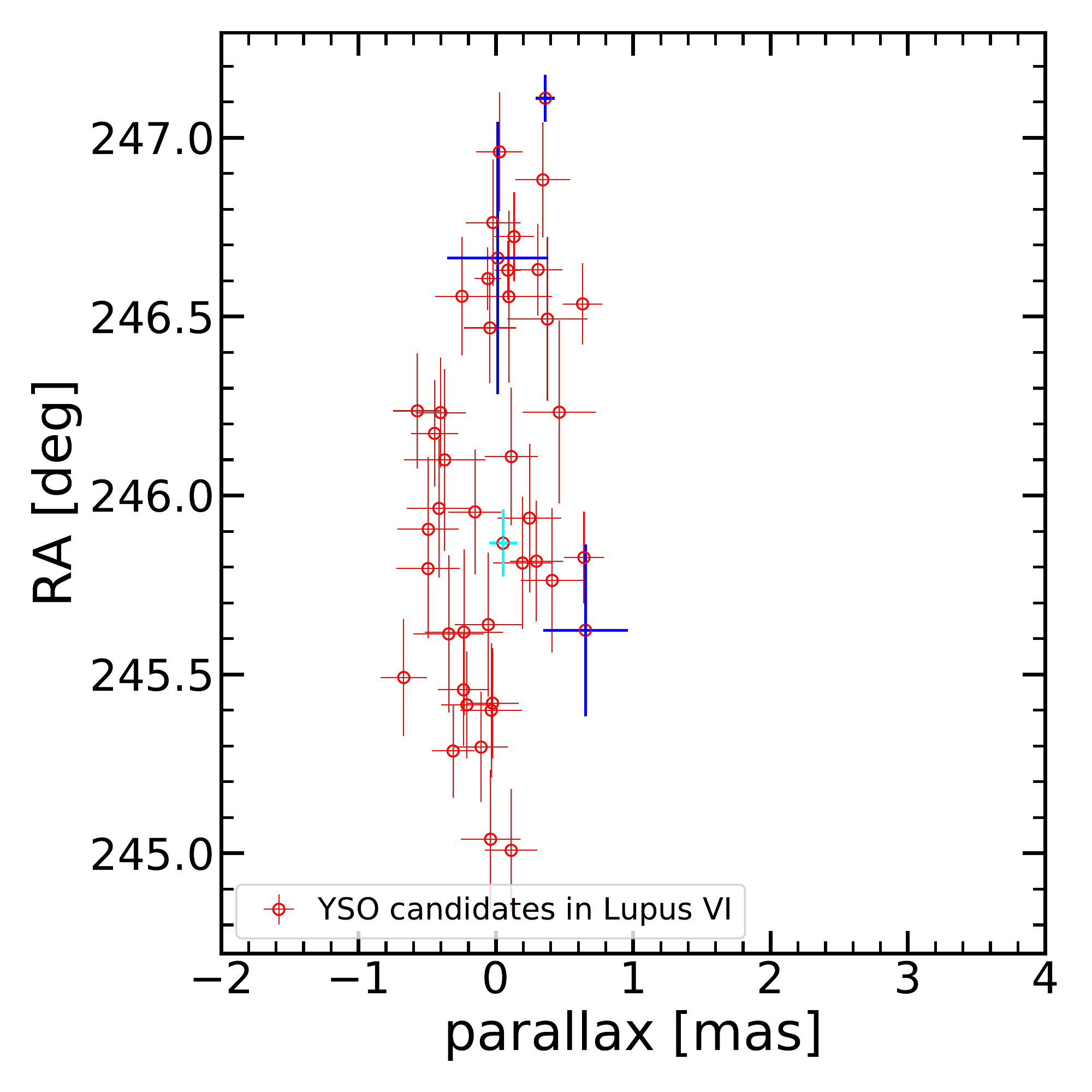}
   \caption{Parallaxes of candidate YSOs as a function of right ascension for objects in the Lupus~V (left) and Lupus~VI (right) region of the sky. Blue crosses indicate candidate Class~II objects, cyan crosses candidate transition disk objects. }
              \label{fig::parra}%
    \end{figure}

Here we explore whether there is any dependence of the measured parallax on the position of the objects in the sky, which could imply that only portions of the clouds contain YSOs. Figure~\ref{fig::pardec} shows that there is no trend of parallaxes with declination, and that the five targets with parallax $>$ 5 are, in fact, located at different declinations within the Lupus~V cloud. Instead, Fig.~\ref{fig::parra} shows that all the objects with parallax $>$5 mas are on the west side of the cloud. In these figures, the YSO candidates classified as Class~II objects are highlighted with a blue cross. 
%

\begin{figure*}
   \centering
  \includegraphics[scale=0.45]{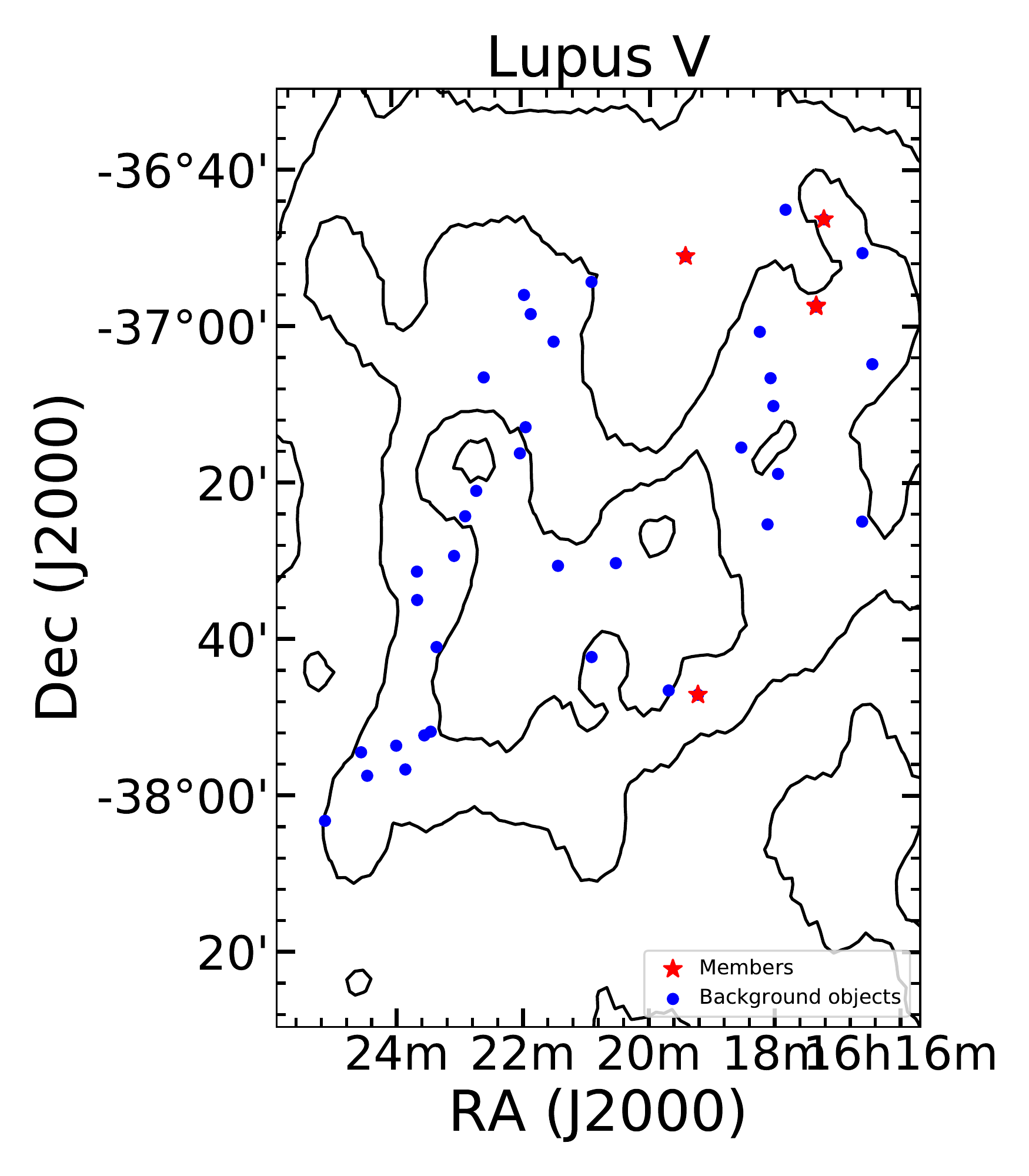}
   \caption{Extinction map from IRAS and COBE/DIRBE \citep{schlegel98}. Contours correspond to these values of E(B-V): 0.66, 0.91, 1.17, 1.42, 1.68. The symbols of two of the members overlap. }
              \label{fig::mapv}%
    \end{figure*}

We show the location of the candidate YSOs and the actual members in Fig.~\ref{fig::mapv} superimposed on the extinction map by \citet{schlegel98} \footnote{Obtained at https://irsa.ipac.caltech.edu/applications/DUST/}. The members are found to be in the lower extinction part of the cloud, and mainly on the west side of it, thus in the direction of the Lupus~III cloud.

\begin{thebibliography}{}

\bibitem[Alcal{\'a} et al.(2014)]{alcala14} Alcal{\'a}, J.~M., Natta, A., Manara, C.~F., et al.\ 2014, \aap, 561, A2 

\bibitem[Alcal{\'a} et al.(2017)]{alcala17} Alcal{\'a}, J.~M., Manara, C.~F., Natta, A., et al.\ 2017, \aap, 600, A20 

\bibitem[Comer{\'o}n(2008)]{comeron08} Comer{\'o}n, F.\ 2008, Handbook of Star Forming Regions, Volume II, 5, 295 

\bibitem[Comer{\'o}n et al.(2009)]{comeron09} Comer{\'o}n, F., Spezzi, L., \& L{\'o}pez Mart{\'{\i}}, B.\ 2009, \aap, 500, 1045 

\bibitem[Cox et al.(2016)]{cox16} Cox, N.~L.~J., Arzoumanian, D., Andr{\'e}, P., et al.\ 2016, \aap, 590, A110 


\bibitem[Dunham et al.(2015)]{dunham15} Dunham, M.~M., Allen, L.~E., Evans, N.~J., II, et al.\ 2015, \apjs, 220, 11 

\bibitem[Evans et al.(2009)]{evans09} Evans, N.~J., II, Dunham, M.~M., J{\o}rgensen, J.~K., et al.\ 2009, \apjs, 181, 321-350 

\bibitem[Facchini et al.(2016)]{facchini16} Facchini, S., Clarke, C.~J., \& Bisbas, T.~G.\ 2016, \mnras, 457, 3593 


\bibitem[Fedele et al.(2010)]{fedele10} Fedele, D., van den Ancker, M.~E., Henning, T., Jayawardhana, R., \& Oliveira, J.~M.\ 2010, \aap, 510, A72 

\bibitem[Frasca et al.(2017)]{frasca17} Frasca, A., Biazzo, K., Alcal{\'a}, J.~M., et al.\ 2017, \aap, 602, A33 

\bibitem[Gaia Collaboration et al.(2016)]{gaia} Gaia Collaboration, Prusti, T., de Bruijne, J.~H.~J., et al.\ 2016, \aap, 595, A1 

\bibitem[Gaia Collaboration et al.(2018)]{gaiadr2} Gaia Collaboration, Brown, A.~G.~A., Vallenari, A., et al.\ 2018, arXiv:1804.09365 

\bibitem[Gaia Collaboration et al.(2018)]{gaiadr2_hdr} Gaia Collaboration, Babusiaux, C., van Leeuwen, F., et al.\ 2018, arXiv:1804.09378 

\bibitem[Gregorio Hetem et al.(1988)]{musca88} Gregorio Hetem, J.~C., Sanzovo, G.~C., \& Lepine, J.~R.~D.\ 1988, \aaps, 76, 347 

\bibitem[Haisch et al.(2001)]{haisch01} Haisch, K.~E., Jr., Lada, E.~A., \& Lada, C.~J.\ 2001, \apjl, 553, L153 

\bibitem[Hern{\'a}ndez et al.(2007)]{hernandez07} Hern{\'a}ndez, J., Hartmann, L., Megeath, T., et al.\ 2007, \apj, 662, 1067 

\bibitem[Lada, et al.(2013)]{lada13} Lada, C.~J., Lombardi, M., Roman-Zuniga, C., et al.\ 2013, \apj, 778, 133.


\bibitem[Lindegren et al.(2018)]{gaiadr2astrometry} Lindegren, L., Hernandez, J., Bombrun, A., et al.\ 2018, arXiv:1804.09366 

\bibitem[L{\'o}pez Mart{\'\i}, et al.(2011)]{lopez-marti11} L{\'o}pez Mart{\'\i}, B., Jim{\'e}nez-Esteban, F. \& Solano, E.\ 2011, \aap, 529, A108.

\bibitem[Luri et al.(2018)]{gaiadr2parllax} Luri, X., Brown, A.~G.~A., Sarro, L.~M., et al.\ 2018, arXiv:1804.09376 

\bibitem[Marigo et al.(2017)]{parsec} Marigo, P., Girardi, L., Bressan, A., et al.\ 2017, \apj, 835, 77 

\bibitem[Mer{\'{\i}}n et al.(2008)]{merin08} Mer{\'{\i}}n, B., J{\o}rgensen, J., Spezzi, L., et al.\ 2008, \apjs, 177, 551-583 
\bibitem[Oliveira et al.(2009)]{oliveira09} Oliveira, I., Mer{\'{\i}}n, B., Pontoppidan, K.~M., et al.\ 2009, \apj, 691, 672 

\bibitem[Prato et al.(2008)]{prato08} Prato, L., Rice, E.~L., \& Dame, T.~M.\ 2008, Handbook of Star Forming Regions, Volume I, 4, 18 

\bibitem[Robin et al.(2012)]{robin12} Robin, A.~C., Luri, X., Reyl{\'e}, C., et al.\ 2012, \aap, 543, A100 

\bibitem[Romero et al.(2012)]{romero12} Romero, G.~A., Schreiber, M.~R., Cieza, L.~A., et al.\ 2012, \apj, 749, 79 

\bibitem[Rosotti \& Clarke(2018)]{rosotti18} Rosotti, G.~P., \& Clarke, C.~J.\ 2018, \mnras, 473, 5630 

\bibitem[Schlegel et al.(1998)]{schlegel98} Schlegel, D.~J., Finkbeiner, D.~P., \& Davis, M.\ 1998, \apj, 500, 525 

\bibitem[Soderblom et al.(2014)]{soderblom14} Soderblom, D.~R., Hillenbrand, L.~A., Jeffries, R.~D., Mamajek, E.~E., \& Naylor, T.\ 2014, Protostars and Planets VI, 219 

\bibitem[Spezzi et al.(2011)]{spezzi11} Spezzi, L., Vernazza, P., Mer{\'{\i}}n, B., et al.\ 2011, \apj, 730, 65 

\bibitem[Tachihara et al.(2001)]{tachihara01} Tachihara, K., Toyoda, S., Onishi, T., et al.\ 2001, \pasj, 53, 1081 


\end{thebibliography}
\end{document}